\documentclass[twocolumn,showpacs,preprintnumbers,amsmath,amssymb]{revtex4}
\usepackage{graphicx}

\begin{document}

\title{\emph{0.7-anomaly} and magnetotransport of disordered\\ quantum wires}

\author{M.~Czapkiewicz}
\author{P.~Zagrajek}
\author{J.~Wr\'obel}
\author{G.~Grabecki}
\author{K.~Fronc}
\author{T.~Dietl}
\altaffiliation[Also at ]{
Instytut Fizyki Teoretycznej, Uniwersytet Warszawski, Ho\.za
69, PL 00-681 Warszawa, Poland}
\altaffiliation{ERATO Semiconductor Spintronics Project, Japan Science Technology Agency, 1-18 Kitamemachi, Aoba-ku Sendai, 980-0023, Japan}

\affiliation{Instytut Fizyki Polskiej Akademii Nauk, al. Lotnik\'ow 32/46,
PL 02-668 Warszawa, Poland 
}

\author{Y.~Ono}
\author{S.~Matsuzaka}
\altaffiliation[Also at ]{ERATO Semiconductor Spintronics Project, Japan Science and Technology Agency, 1-18 Kitamemachi, Aoba-ku Sendai, 980-0023, Japan}
\author{H.~Ohno}
\altaffiliation[Also at ]{ERATO Semiconductor Spintronics Project, Japan Science and Technology Agency, 1-18 Kitamemachi, Aoba-ku Sendai, 980-0023, Japan}

\affiliation{Laboratory for Nanoelectronics and Spintronics, Research Institute of Electrical Communication, Tohoku University, Katahira 2-1-1, Aoba-ku, Sendai 980-8577, Japan}

\begin{abstract}
The unexpected "0.7" plateau of conductance quantisation
is usually observed for ballistic one-dimensional devices. In this
work we study a quasi-ballistic quantum wire, for which the
disorder induced backscattering reduces the conductance quantisation
steps. We find that the transmission probability resonances
\emph{coexist} with the anomalous plateau. The studies of these
resonances as a function of the in-plane magnetic field and electron
density point to the presence of spin polarisation at low carrier
concentrations and constitute a method  for the determination of the
effective \emph{g-}factor suitable for disordered quantum wires.
\end{abstract}


\maketitle

It is expected that quantum point contacts (QPC) and quantum wires
(QW) will act as active components of future nano-electronic devices
and circuits. Therefore, the renewed interest in transport and spin
properties of one-dimensional (1D) systems recently takes place in
the mesoscopic physics community. In those studies, special
attention is directed towards the long standing problem of quantum
transport -- the so called ``\emph{0.7 anomaly}'' \cite{1996thomas}
most often, but not exclusively, observed for devices fabricated on
modulation doped GaAs/AlGaAs heterostructures. Usually, anomalous
behavior is observed in transport data as a ``kink'' on the
conductance $G$ vs. the device width curve, occurring for the low
carrier densities, when $G \sim 0.7 \times 2e^2/h$, here $-e$ is the
electron charge and $h$ is the Planck constant. The origin of this
effect is currently under active debate since this anomaly seems to
be an universal, but still unexplained feature of one-dimensional
mesoscopic transport. Experimentally, the magnetic field dependence
of the additional plateau is common for all studied systems -- by
applying a parallel in-plane field the 0.7 feature evolves gradually
towards $0.5 \times 2e^2/h$ conductance step, when only one
spin-polarised level is occupied
\cite{1996thomas,1998thomas,2002cronenwett,2003graham,2005chou,2007koop}.
Therefore, it has been suggested that such an anomalous plateau is
due to spontaneous spin polarisation of one-dimensional electron
liquid, caused by  exchange interactions among carriers in the
constricted geometry of the device \cite{1996wang,1998thomas}. If it
is so, the 1D systems may be used as an efficient spin filter with
possible practical applications. This point of view is supported by
magnetic focusing data obtained for the p-type device, which reveal
the static spin polarisation of holes transmitted through the
constriction \cite{2006rokhinson}. Furthermore, recent shot-noise
measurements carried out for n-type QPC \cite{2006dicarlo} show that
distinct transport channels exist at $G < 2e^2/h=G_0$, presumably
related to spin, exhibiting quite different transmission
probabilities.

Many experiments, however, bring out rather contradictory
observations regarding the temperature dependence of the additional
plateau. Already Thomas and co-workers \cite{1996thomas} revealed
that the 0.7 ``kink'' disappears when temperature is lowered,
typically below few hundreds milikelvins. Such unusual
low-temperature behaviour is accompanied by the zero-bias peak in
the differential conductance, which is typical for the Kondo effect
for quantum dots \cite{2002cronenwett,2002meir}. Nevertheless, the
Kondo-type features appear to be typical for the  point contacts
only, {\em i.e.} for devices for which $L/W \sim 1$, where $L$ and
$W$ are physical length and width of the conducting channel,
respectively \cite{2000kristensen,2007koop,2007graham}. On the
contrary, for longer \emph{quantum wires}, when $L/W \gg 1$,  the
0.7 anomaly becomes even more pronounced, when the temperature is
\emph{lowered} and zero bias anomaly is not observed
\cite{2001reilly,2004picciotto,2005knop}. Additionally, the
anomalous plateau does not always occur at $G=0.7$ (in $G_0$ units),
there is some evidence that the value decreases with the length of
the wire approaching the ``quantized'' value $\sim 0.5$
\cite{2001reilly, 2005knop}. This may suggest that the additional
plateau is generic for $1D$ systems. The Kondo physics shows up
only, when the source-drain distance  is reduced towards zero and
the  confining potential forms a smooth saddle point in the $2D$
landscape.

Usually, the ``$0.7-0.5$'' anomaly is observed for ballistic point
contacts and wires, {\em i.e.}, when $L\ll\ell$, where $\ell$ is the
mean free path of electrons (or holes). As a consequence, in the
analysis of the possible origin of this anomaly, the role of
disorder is ignored. The aim of our work is to study the 0.7 plateau
for quasi-ballistic devices and to detect the fate of the mysterious
conductance ``kink'', when the back-scattering events operate within
the $1D$ channel. For this purpose we have fabricated quantum wires
with $L\sim\ell$ from a wafer containing a modulation doped
AlGaAs/GaAs/AlGaAs quantum well. We show here that the ``0.7
anomaly'' is very robust against the disorder. The analysis of
linear and non-linear transport measurements enable us to observe
the characteristic anomalous plateau for devices with $L/W \approx
20$  in spite that the disorder reduces the magnitude of the overall
conductance and leads to the appearance of transmission resonances.
Actually, as we show here the presence of the resonances allows one
to determine the evolution of spin polarisation with the in-plane
magnetic field and electron density. Employing this method we
determine the electron Land\'e factor $g$ at relatively high carrier
concentrations. At the same time, we find that splitting of the
resonances is absent when the conductance is reduced below
$G\lesssim 0.8$. This observation is in agreement with the
explanations of the 0.7 anomaly in terms of spontaneous spin
polarisation of low density 1D carrier liquid.

The four-terminal quantum wires are patterned of  an MBE-grown 
(by the Veeco GEN-II system)
20~nm AlGaAs/GaAs/AlGaAs:Si quantum well located at 101~nm below the
surface. The 60~nm top barrier  results in the electron
concentration $n_{2D}= 1.8\times 10^{11}$~cm$^{-2}$ and carrier
mobility $\mu=2.45\times 10^5$~cm$^2$/Vs as measured in dark at
$T=2.8$~K. The wires of length $L=0.6$~$\mu$m and lithographic width
$W_{lith}=0.4$~$\mu$m are patterned by \emph{e}-beam lithography and
shallow-etching techniques. The physical width of the wires is
controlled by means of the top metal gate which is evaporated over
the entire device. The differential conductance $G=dI_{sd}/dV_{sd}$
measurements are conducted in a He-4 cryostat and He-3/He-4 dilution
refrigerator by employing a standard low-frequency lock-in technique
with $ac$ voltage excitation of 10 $\mu$V. The source-drain voltage
$V_{sd}$ and current $I_{sd}$ are measured by employing battery
powered, low-noise $dc$ amplifiers.

\begin{figure}
\includegraphics{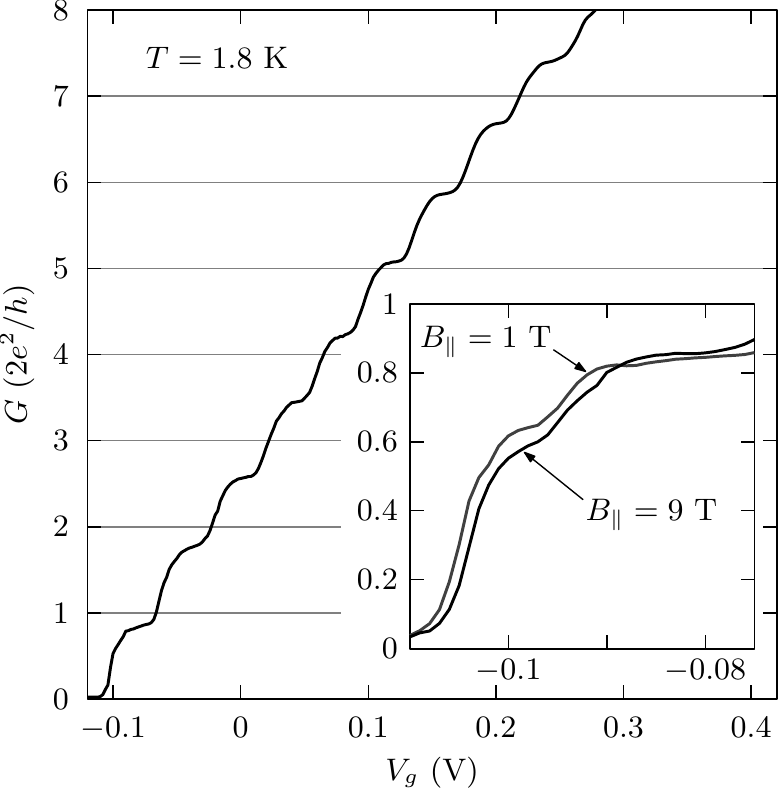} 
\caption{\label{fig.1} Conductance versus gate voltage. Evenly
spaced quantised conductance steps
are visible up to $N=10$, the average step height is $\approx 0.81$
in the $G_0$ units. Inset shows the evolution of the anomalous step
($G \approx 0.6G_0$) - when an external magnetic field of 9~T is applied
parallel to the wire plane.} 
\end{figure}

\begin{figure}
\includegraphics{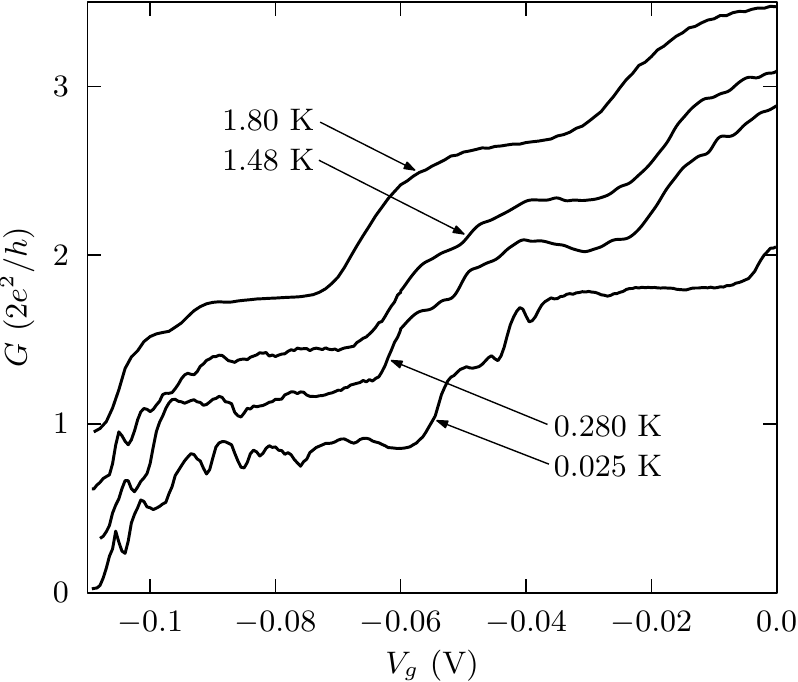} \caption{\label{fig.2} Zero-bias conductance as a function of
the gate voltage at different temperatures, measured after separate
warm-up/cool-down cycles. For clarity curves taken at higher temperatures
are shifted upwards and also moved slightly along the $V_g$-axis, in order
to eliminate small changes in the pinch-off voltage.}
\end{figure}

Figure~\ref{fig.1} presents the conductance $G$ as a function of the
gate voltage $V_g$ obtained at zero $dc$ source-drain bias. The
quantized conductance plateaux are observed, which correspond to the
successive population/depopulation of the 1D electric subbands.
However, the height of particular steps deviates from the quantised
values, $NG_0$, where $N$ is an integer corresponding to the number
of occupied subbands. We find from our four terminal measurements
that the perfect quantisation cannot be recovered by subtracting a
series resistance. Some authors have considered the influence of
electron-electron interactions in 1D systems on conductance
quantisation but the prevailing view is that a reduction in plateaux
heights results from elastic backscattering, occurring within the
disordered quantum channel \cite{1990kramer, 1991tekman}. We
conclude, therefore, that in our wire $G$ is diminished by
backscattering, $G=(2e^2/h)N\mathcal{T}$, where $\mathcal{T}\approx
0.81$ is the transmission probability similar for each channel up to
$N =10$.

However, the disorder not only reduces uniformly the conductance
magnitude but also results in the appearance of conductance
resonances at sufficiently low temperatures, which originate from
interference of scattering amplitudes. The coexistence of these two
effects was already observed experimentally in quasi-ballistic
quantum wires \cite{1992wrobel,1998wrobel,1999liang} and is also
present for the device studied here. Figure~\ref{fig.2} shows
conductance measured at various temperature in the He-4 cryostat and
the dilution refrigerator. With lowering temperature, the
Ramsauer-like resonances clearly show up. This pattern undergoes a
change when the sample is warmed-up and cooled-down again, as
scattering centres move or change their charge state
\cite{1991tekman}. We conclude, therefore, that our sample is indeed
in the quasi-ballistic regime,  $L \sim \ell$.

\begin{figure}
\includegraphics{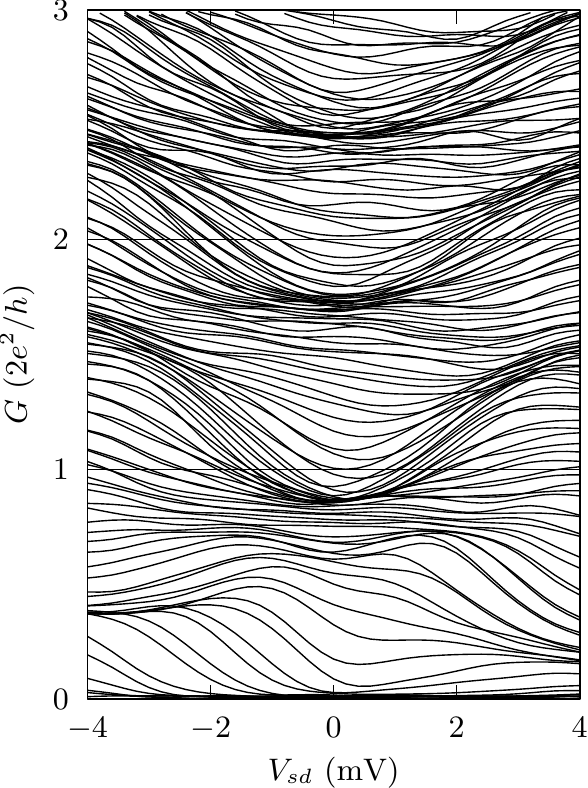} \caption{  \label{fig.3} Differential conductance
versus  source-drain bias voltage $V_{sd}$ at 1.8~K. Each
trace corresponds to the fixed gate voltage $V_g$, which changes
from $0$ to $-0.13$~V with $1$~mV step.
}
\end{figure}

\begin{figure}
\includegraphics{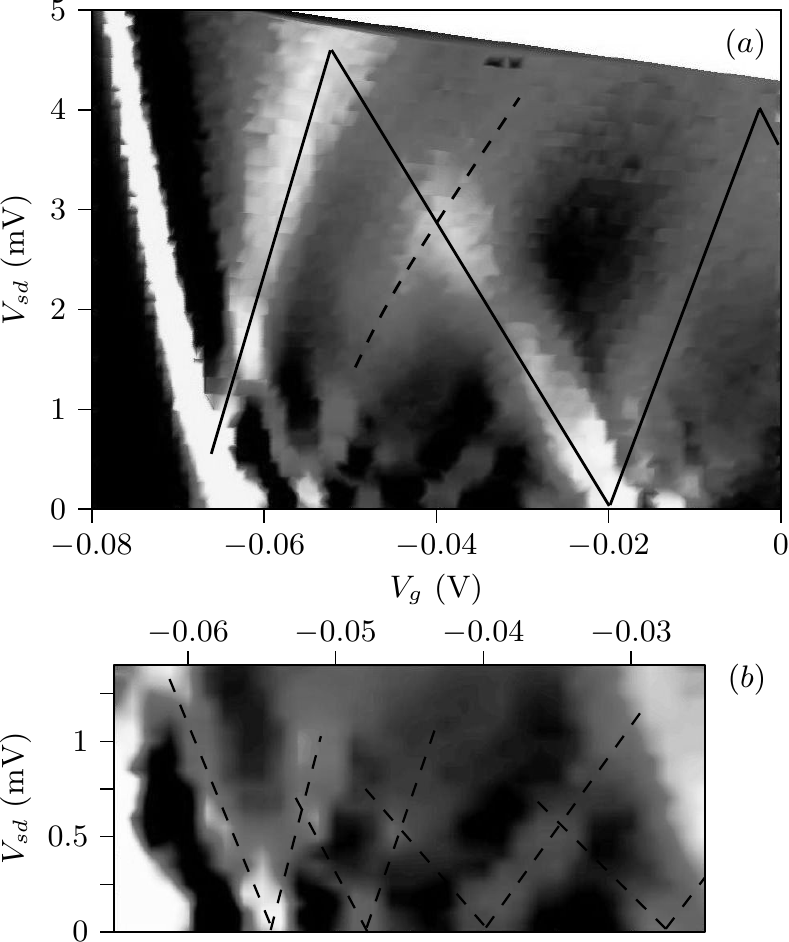} \caption{\label{fig.4} (a) Intensity plot of the
transconductance $dG/dV_g$ versus the gate voltage $V_g$ and
source-drain bias voltage $V_{sd}$ at 25~mK. Bright areas
correspond to large transconductance, {\em i.e.}, to the transition regions
between plateaux (or transmission resonance peaks). Transition points
between plateaux
$N \rightarrow N+1$ splits with the bias because for left- and
right-moving electrons it occurs at different gate voltages, as
marked by solid black lines. The dashed line labels the range where
anomalous plateau changes over to normal one. (b) The enlarged part
corresponding to the $N=1$ plateau with superimposed interference
peaks, dashed lines mark the "sub-diamond" structure, which is
related to such conductance resonances at $V_{sd} \lesssim 1$ mV.}
\end{figure}

The important aspect of our findings is that despite the strong
effects of disorder, the anomalous "0.7" conductance step is visible
on the the $N=1$ plateau. In our case it assumes the value of
$G\approx 0.6G_0$. As expected, when the in-plane magnetic field
$B_{\parallel}$ is applied, the height of this additional plateau
decreases, as shown in the inset to fig.~\ref{fig.1}.

A question arises, whether the conductance step in question is not a
resonance peak persisting up 2~K. In order to address this issue, we
have carried out dynamic conductance measurements as a function of
source-drain bias \cite{1991patel}, as it is known that the "0.7"
plateau becomes even wider for non-zero bias and can survive up to
$V_{sd} \approx 10$~meV \cite{2000kristensen,2004picciotto}.
Figure~\ref{fig.3} shows the evolution of differential conductance
with the bias and gate voltage. Quantized plateaux are visible as
collections of lines, which for $V_{sd} \gtrsim 4$~meV group again
to form a ``half-integer'' plateaux, when the  left- and right-
moving electrons occupy different consecutive energy levels. For $G
< 0.8$ the aggregation of lines is identified as the ``0.6''
anomaly, discernible also for non-zero biases. The distinction
between the additional plateau and any interference effect is
evidenced in Fig~\ref{fig.4}, where the transconductance $dG/dV_g$
at $T=0.025$~K is shown as a function of gate voltage and bias.
Together with the distinctively diamond-shaped regions, bordered by
high transconductance (bright) areas, also the additional bright
strip which crosses the borderline between integer and half-integer
conductance diamonds is clearly visible. According to previous
studies such characteristic high transconductance line forms when
the ``0.7-0.5'' plateau evolves toward the normal one and it is observed
up to biases comparable with the energy gap between $1D$ levels.
Therefore we conclude that the ``0.6'' anomaly is present also at
low temperatures, provided $V_{sd}>~1$~mV. In Fig~\ref{fig.4}(a) the
anomalous plateau is visible as a darker region on the left side of
dashed curve. At low source-drain voltages, however, the anomaly is
masked by transmission resonances which show-up due to the presence
of disorder. As a result, the characteristic high transconductance
line is replaced by the ``sub-diamond'' pattern, which forms because
the interference pattern splits in energy. Distinct branches for
electrons injected from left and right are marked with dashed lines
in Fig~\ref{fig.4}(b).

Recently, Bielejec \cite{2003bielejec} {\em et.~al.} have reported a
reduction of the conductance step (down to $0.84$) \emph{and} the
presence of "0.7" anomaly for a long and very high mobility
split-gate device.  However, no Ramsauer-like transmission
resonances were found at low temperatures. For the data reported
here, we can observe a stronger reduction of conductance, the "0.6"
anomaly, and transmission resonances. It seems that the wires
patterned by shallow etching are well suited for such studies
because the 1D confining potential is much stronger comparing to
split-gate devices \cite{2000kristensen,2005knop}. From the
source-drain bias spectroscopy we obtained $E_2-E_1=4.6 \pm
0.2$~meV, where $E_N$ is the $N$-th level energy. Assuming a
parabolic confinement we estimate the physical width of the device
as $W \approx 0.03$~$\mu$m (for $V_g=0$). As expected for such a
high aspect ratio, we have not observed any zero bias anomaly at
25~mK, which in Fig~\ref{fig.4}(a) should manifest itself as an
additional white spot for $V_{sd}\rightarrow 0$ and $V_g\lesssim
-0.068$~V. The data for low biases are, however, obscured by the
presence of transmission resonances, so this conclusion
requires further studies.

\begin{figure}
\includegraphics{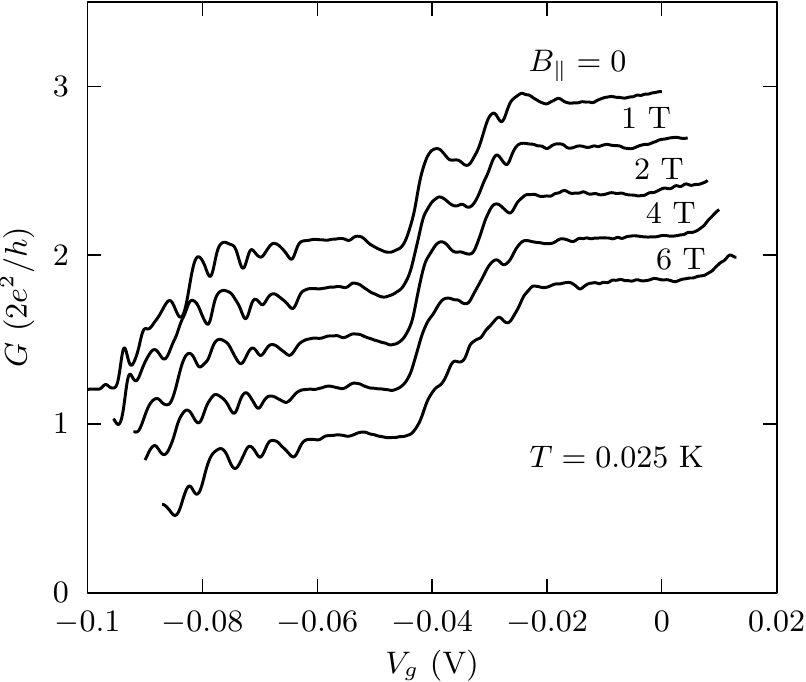} \caption{\label{fig.5} Zero bias conductance as a function of
the gate voltage at 25~mK for different in-plane magnetic field
strengths. Except for the data for $B=6$~T curves are shifted
relative to each other and also slightly moved along $V_g$-axis to
line-up positions of conductance peaks. For this particular
cool-down cycle the absolute value of pinch-off voltage was smaller
as compared to data presented in Fig~\ref{fig.2}} 
\end{figure}

\begin{figure}
\includegraphics{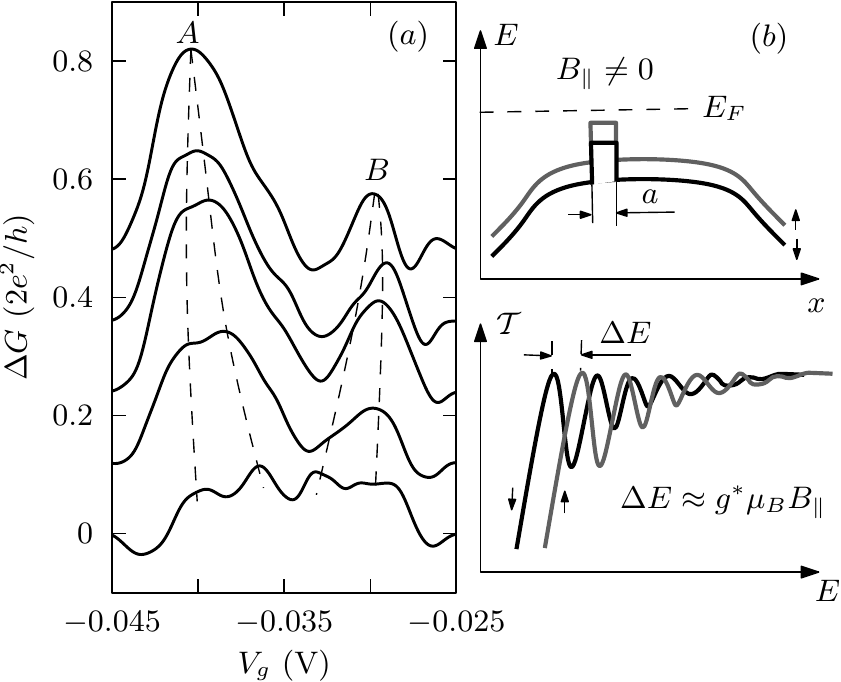} 
\caption{\label{fig.6} (a) Fluctuating part of the
conductance which is observed when the $N=2$ level starts to be
populated by sweeping the gate voltage $V_g$. Except for data at
$B=6$~Tcurves are shifted upwards for clarity, top curve corresponds
to $B_\parallel = 0$, below data for $1$, $2$, $4$ and $6$~T are
shown. Dashed lines are guide for the eye and indicate the magnetic
field evolution of transmission peaks marked $A$ and $B$. (b) Upper
part shows schematically the smooth potential along the wire with
superimposed short range ($a<L$) barrier as experienced by spin-up
and spin-down electrons when $B_\parallel \neq 0$. Lower part
explains the resulting splitting of Ramsauer resonances $\Delta E$
which occurs when Fermi energy $E_F$ approaches the barrier height.}
\end{figure}

If the interference pattern splits when left- and right-moving
electrons start to differ in energy, we may ask if the Zeeman
splitting induced by the external field, will lead to a similar
effect. Actually, the influence of spin splitting upon mesoscopic
conductance fluctuations was observed for diffusive transport in
quantum wires of diluted magnetic semiconductor (Cd,Mn)Te:In
\cite{1995jaroszynski}. It was found that the correlation field
$B_c$ of the universal conductance fluctuations scales with the
$s-d$ exchange spin splitting of the conduction band. Therefore we
expect that in non-magnetic materials the interference pattern will
also split at a sufficiently high in-plane magnetic field - when the
Zeeman energy exceeds the separation of fringes. Relevant data for
our sample is presented in Fig.~\ref{fig.5} where the evolution with
the magnetic field of the first two conductance steps is shown.

As seen, a different behaviour for $N=1$ and $N=2$ plateaux is
observed. In particular, in the region where $G < 0.8$ transmission
resonances do not split -- the amplitude of fluctuations and the
number of peaks is approximately constant. On the contrary, when the
$0.8 < G < 1.6$ interference pattern changes, as it is shown more
clearly in Fig.~\ref{fig.6}(a), where only the fluctuating part
$\Delta G$ of the total conductance is shown. The amplitude of the
two most prominent peaks $A$ and $B$ decrease and they clearly split
at $B_\parallel=6$~T. We note that this behaviour is consistent with
the models of the "0.7" anomaly, which assume the presence of
spontaneous spin polarisation at low carrier densities. If only one
spin sublevel is occupied below the $N=1$ plateau, we do not expect
any significant changes of the interference pattern with the
magnetic field, as indeed observed in the experiment. If, however,
both spin channels participate in charge transport, the Zeeman
effect should split the Ramsauer resonances, as explained
schematically in Fig.~\ref{fig.6}(b). Furthermore, such splitting
$\Delta E$ should depend only weakly on the exact shape of the
potential barrier introduced by disorder, so that $\Delta E\approx
|g^*|\mu_B B_\parallel$, where $g^*$ is the effective Land\'e factor
and $\mu_B$ is the Bohr magneton. From Fig.~\ref{fig.6} we obtained
that the splitting $\Delta V_g$, which is more clearly visible for
the peak $A$, approaches 4~meV for $B_\parallel =6$~T. It
corresponds to $\Delta E\approx 0.42$ meV, as can determined with
the application of source-drain bias spectroscopy data. Thus, the
magnetic field evolution of transmission resonances can be applied
to the determination of the $g^*$ factor of 1D systems, which is
expected to be enhanced by the exchange interaction
\cite{1998thomas,2003graham}. For our sample we obtained
$|g^*|=1.2\pm 0.1$ (as compared to 0.44 in bulk GaAs), which is in
very good agreement with recent data for ballistic QPCs
\cite{2007koop}. The advantage of the proposed method is that it can
be applied for long, quasi-ballistic quantum wires. The disadvantage
is, of course, the non-reproducibility of observed interference
pattern, which changes after subsequent warming up and then cooling
down.

In summary, we studied the conductance quantization of
quasi-ballistic ($L\sim\ell$), the large aspect ratio ($L/W \approx
20$) quantum wire. Due to the disorder present in our sample, we
observed reduced conductance steps ($(0.81\pm 0.01)\times G_0$) and
transmission resonances on the onset of quantized plateaux, caused by
the interference of incoming and reflected electron waves. With the
application of source-drain bias spectroscopy, we showed, that the
backscattering of electron waves and reduction of transmission
probability through the sample do not destroy the existence of so
the called 0.7 (in our case 0.6) anomaly. The additional conductance
plateau was clearly observed in the temperature range from 1.8 to
0.025~K. Furthermore, we proposed the single electron effect
(splitting of the interference peaks with in-plane magnetic field)
as a suitable tool for studying the enhancement of electron
\emph{g-}factor in quantum wires which is a collective phenomenon,
observed before for ballistic point contacts.

\acknowledgments
We acknowledge support by the Eurocores/ESF grant SPINTRA
(ERAS-CT-2003-980409) and Semiconductor Spintronics ERATO project by JST.


\end{document}